\documentclass[reprint,
nofootinbib,
 amsmath,amssymb,
aps,
]{revtex4-1}
\usepackage[caption=false]{subfig}
\usepackage{graphicx}
\usepackage{dcolumn}
\usepackage{bm}
\usepackage{hyperref}
\usepackage{booktabs}
\usepackage{amsmath}
\usepackage{float}
\usepackage{caption}
\usepackage{multirow}
\usepackage{textgreek}

\graphicspath{{figures/}}

\begin{document}


\title{Dendritic Growth of a Polymer on a 2D Mesoscale Square Lattice}

\author{Joel Martis\(^1\)}
\email{joelmartis95@gmail.com}
\author{Kaushik Satapathy\(^2\)}
\author{P R Shaina\(^3\)}
\author{C V Krishnamurthy\(^3\)}
\author{Manu Jaiswal\(^3\)}
\email{manu.jaiswal@iitm.ac.in}

\affiliation{%
\vspace{0.25cm} \(^1\)Department of Mechanical Engineering, Indian Institute of Technology Madras, Chennai, India - 600036\vspace{0.25cm}\\ \vspace{0.25cm}
\(^2\)Department of Aerospace Engineering, Indian Institute of Technology Madras, Chennai, India - 600036 \\ 
\(^3\)Department of Physics, Indian Institute of Technology Madras, Chennai, India - 600036
}%

\date{\today}

\begin{abstract}
Dendritic growth patterns exhibiting four-fold anisotropy are observed when polyethylene oxide undergoes phase segregation from a solution phase to a solid phase. When this phase transition occurs on a substrate that has patterns of cross-linked polyethylene oxide resist pillars made using electron beam lithography , the polymer grows in between the patterns giving rise to dendritic growth structures exhibiting eight-fold anisotropy. This paper presents these experimental observations and explains the dendritic growth using principles of minimization of free energy associated with phase change. Numerical simulations are carried out using phase-field modeling, and the results are shown to qualitatively match experimental observations. The simulations reveal that the polymer assumes the anisotropy of the patterned lattice.

\end{abstract}

\maketitle


\section{\label{sec:level1}Introduction}

Patterns exhibiting various forms of symmetry are ubiquitous in nature \cite{paper8} and exist in biological systems \cite{paper10,paper11} as well as non-living systems. Popular examples include living creatures such as starfish (exhibiting five-fold symmetry), and objects such as snowflakes (six-fold symmetry) \cite{paper9}. The scale of such patterns can vary greatly \cite{paper1}, with symmetry being observed in the structure of galaxies such as the M74 spiral galaxy and the complex granular structure of the Sun's surface, to sand dune patterns in deserts, to snowflakes that are a few millimeters in width. One of the patterns seen in nature is the dendritic growth pattern \cite{paper12,paper13}, which is observed in the growth of tree branches, growth of neurons \cite{paper14} in the human brain, solidification of amalgams like Diana's Tree (or the Philosopher's Tree), formation of snowflakes, the solidification of polymers etc. Dendritic patterns are fractals \cite{paper15} which may or may not exhibit directional preference (anisotropy). In systems like the human neuron, there is no definitive anisotropy, whereas in most crystalline dendritic systems, one can observe anisotropic growth.\\

Polyethylene oxide (PEO) \cite{paper16,paper17}, also known as polyethylene glycol, is a polymer that is widely used in laboratory studies.  PEO is well known to be exceptionally soluble in water \cite{israelachvili}. PEO exists in water as a random coil without aggregation \cite{devanand}. The mechanism responsible for solubility is hydrogen bonding between oxygen atoms in PEO and water molecules. What is particularly favorable towards this is that the O-O separation in PEO is quite similar to the corresponding separation in hydrogen-bonded water \cite{hammouda}. PEO can also serve as a resist for electron beam lithography, with water as the developer \cite{shaina}. The chemical formula of PEO is given by H-(O-CH\(_2\)-CH\(_2\))\(_n\)-OH. PEO, when mixed with nucleating agents such as clay powder and with other polymers such as PMMA, has been known to exhibit a wide variety of growth structures like spherulitic growth, seaweed dendritic growth, symmetric dendritic growth, and fractal dendritic growth \cite{paper18, paper19, paper20, paper21, paper22}. These growth structures are usually obtained by crystallizing PEO from a liquid or solution phase. Of interest to this paper is the symmetric dendritic growth of PEO which exhibits four-fold anisotropy. These growth patterns are obtained by spin-coating PEO on a substrate (in our case, a SiO\(_2\)/Si wafer) and subjecting the system to ambient relative humidity above 50\%, above a certain temperature. When the PEO coated substrate is lithographically patterned and exposed to ambient humidity above a certain temperature, patterns which grow in between the pillars and exhibit eight-fold anisotropy are formed. Temperature controlled dissolution and resegregation of PEO gives rise to the growth structures.\\

The lower critical solution temperature (LCST) is the temperature below which the polymer is completely miscible in the solution and exists as a single phase. Similarly the upper critical solution temperature (UCST) is also defined as the temperature bound above which the single phase exists. For PEO in water, both these temperature values are outside the typical range of interest, 0-100 \(^\textrm{o}\)C interval \cite{seuring}. However, a novel mechanism involving monomer-monomer and monomer-water interactions has been reported wherein PEO crystallizes in water around 66 \(^\textrm{o}\)C in the weight fraction range 0.5 to 1 \cite{bekiranov}. In addition to this, it is reported that the presence of tiny amounts of organic impurities or dust particles can shift the thermodynamics to cause association of PEO chains in the aqueous solution and this can result in temperature dependent phase segregation for solutions which are more dilute \cite{devanand}.\\

Numerical modeling of phase segregation driven by temperature using phase field modeling validates this claim. It is observed via simulations that pillar separation plays a major role in deciding the anisotropy of the final growth structure. Numerical simulations also reveal that the anisotropy of the growth depends on the configuration and anisotropy of the pillar pattern, and not on the polymer's intrinsic anisotropy.\\

The experiment and its results are described in detail in section II. In section III, numerical modeling of the system is discussed. Section IV presents the results of numerical simulations and a discussion comparing these results with experiment. Section IV sums up the paper with conclusions and scope for future research.\\

\section{\label{sec:level1}EXPERIMENT}
\subsection{\label{sec:level3}Method}

PEO thin films of thickness 10-15 nm are prepared by spin-coating 1 wt. \% aqueous solution on SiO\(_{2}\)/Si wafers. A square array of PEO pillars of diameter 200 nm is created on the wafer using electron beam lithography, with an optimized dose of 50 \textmu C/cm\(^2\) \cite{shaina} . The square array of PEO pillars is spread over an area of 150 x 150 \(\mu\)m on the wafer, and the separation between the edges of adjacent pillars is 1 \textmu m. Following electron beam lithography, the wafer is either (a) immersed in water for about 60 seconds or (b) exposed to ambient humidity for a few hours, to dissolve the unexposed polymer. Wafers that are dipped in water are then dried by placing them on a hot plate at 80 \(^\textrm{o}\)C for five minutes. They are then observed under a Scanning Electron Microscope (SEM).\\

\begin{figure*}
	\centering
	\includegraphics[width=1\textwidth]{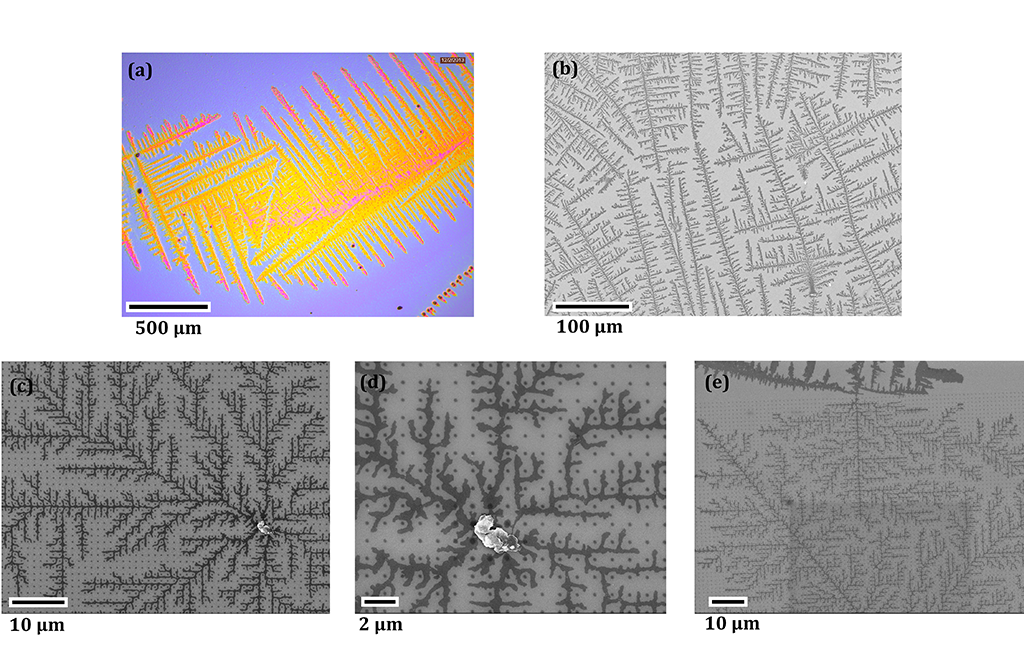}
	\label{fig1}
	\caption{(a) Optical image of a bare wafer with dendritic growth, (b) SEM image of a bare wafer with dendritic growth patterns, (c) SEM image of a patterned wafer with dendritic growth patterns, (d) Magnified SEM image of the dendritic growth on the patterned wafer, (e) SEM image of a patterned wafer with dendritic growth patterns showing the transition between growth inside and outside the pattern.}
\end{figure*}

\subsection{\label{sec:level3}Results and Discussion}
 During electron beam lithography, the electron beam of energy 10-30 kV causes cross-linking of PEO polymer chains \cite{shaina}, thus increasing their molecular weight. When the PEO films are dipped in water for developing the pattern, the pillars created during electron beam lithography do not dissolve due to their high molecular weight \cite{shaina}. \\
 
Let us examine the PEO film coated on the substrate before it is exposed to lithography and patterns are created. Spin-coated  films of PEO of homogeneous thickness were left in ambient conditions (RH \(>\) 50 \%), for extended periods of time (about few weeks). An optical image of such a film is shown in Figure 1(a). It can be seen that some dendritic growth patterns are observed in these areas and they are of characteristic four fold anisotropy. The formation of these dendritic structures is well known \cite{paper18,paper19,paper20}. It is understood that the PEO film is locally dissolved in the ambient moisture, though the dissolution in surface water should typically lead to higher weight concentration of PEO. These are the conditions known to favor re-segregation of PEO to solid form at temperatures more than 30 \(^\textrm{o}\)C below the LCST of PEO \cite{bekiranov}. This segregation leads to the formation of characteristic dendritic structures. Figure 1(b) shows an SEM image of dendritic growth structures on a bare substrate. The features of the growth patterns are better resolved in the SEM images as compared to optical images. Some salient features observed in the growth are (1) The dendritic growth is self-avoiding, and (2) The growth exhibits four fold anisotropy. Figure 1(c) shows an SEM image of dendritic growth on a patterned substrate. For the patterned substrates, the role of ambient moisture is replaced by the direct treatment of the sample with water solution, which is needed to develop the PEO pillars. The PEO that grows in dendritic structures is the residual unexposed PEO when the surface water is removed by heating around 80 \(^\textrm{o}\)C. It is seen the growth avoids the pillars within the patterned region in addition to displaying self-avoidance. Further, the dendritic growth structure inside the patterned region seems to exhibit eight fold anisotropy. Details of the nature of growth are seen in the magnified SEM image Figure 1(d). Figure 1(e) shows an SEM image of a dendritic growth structure occurring both within the patterned region and outside it. It can be inferred from the image that the growth occurring inside and outside the patterned region is the same, meaning that the growth occurring inside the patterned region is a result of the growth in the bare region plus the constraints imposed by the pillars. It is also seen that the four-fold anisotropic growth that occurs outside the patterned region seems to exhibit eight fold anisotropy within the patterned region.\\

From the discussion in this previous section, it is clear that a theory used to model the growth should account for phase segregation controlled by a characteristic temperature. It should also explain self-avoidance and avoidance of pillars by the polymer. Finally, the theory should explain the resulting anisotropy of the dendritic growth in the presence and absence of a 2D mesoscale lattice.

\section{\label{sec:level1}PHASE-FIELD MODELING}

In this section, we discuss how the system is modeled using the principle of free energy minimization associated with phase change. The approach used is very general and is applicable to any system involving phase segregation and dendritic growth. This means that temperature dependent dissolution and phase segregation give rise to dendritic growth structures. The free energy is modeled such that the system exhibits intrinsic four-fold anisotropy, thus giving rise to dendritic growth structures as seen in the experiment. To model the patterned substrate, pillars are modeled as regions having a higher `effective temperature' than the surrounding region, thus making growth unfavorable, or creating a sort of repulsion. In reality, the `effective temperature' can be a proxy for electronic repulsion or some other forces involved. A phase field model is to numerically simulate the system since it does not require explicit interface tracking. Since many of the microscopic material constants required to carry out numerical simulations are typically not known, sample values are used with the aim of qualitatively comparing the results of numerical simulations with experimental observations discussed in the previous section.\\

Section A discusses details of the theory used to model the system, such as the exact form of free energy used, differential equations to be solved and so on. Section B discusses the details of numerical simulations. Section C discusses the results obtained from numerical simulations.
\vspace{-1cm}
\vspace{0.5cm}
\subsection{\label{sec:level2}Theory}

\begin{figure}[h!]
	\centering
	\hspace{0.5cm}
	\includegraphics[width=0.5\textwidth]{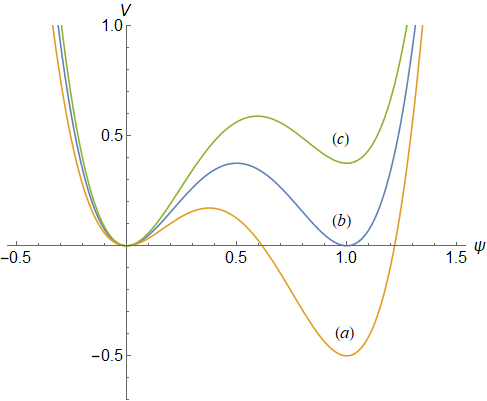}
	\caption{\(V\) v/s \(\psi\) plot for (a) \(k=0.125\), (b) \(k=0\) and (c) \(k=-0.1125\). Note that \(A=6\) in all three cases.}
	\label{figure6}
\end{figure}

The system is physically modeled in two dimensions using phase-field modeling \cite{paper2,paper3,paper4,paper5,paper6}. In particular, we closely follow the approach used by Ryo Kobayashi \cite{paper7}. This involves defining two scalar fields, namely the phase field \(\psi(x,y,t)\) and the temperature field \(T(x,y,t)\), where \(x\) and \(y\) denote spatial coordinates and \(t\) is the time coordinate. The phase field variable \(\psi\) (also called the order parameter) denotes the fraction of solid phase at a point, and hence, ranges in value from 0 to 1, with 0 signifying the solution phase and 1 signifying the solid phase. The advantage of using such a model is that explicit tracking of the interface is not required, and the interface can be traced over a finite region over which \(\nabla\psi\ne\)0. The free energy at any point is a function of \(\psi\) and \(T\), and is expressed as 

\begin{equation}
\label{equation1}
F = \iint (\frac{1}{2} \xi^2 |\nabla\psi|^2 + V(\psi,k)) dxdy
\end{equation}

where \(\xi\) is a microscopic interaction length, \(V\) is the free energy of the bulk material and \(k\) is a parameter that depends on temperature as \(k=\frac{\mu}{\pi}\tan^{-1}(\lambda(T_{c}-T))\). \(T_{c}\) is the critical temperature and \(\mu\) and \(\lambda\) are constants. The exact form of \(V\) is not known for the system in the experiment, so it is assumed that it satisfies a Ginzburg-Landau type of double well potential. The specific form of \(V\) used in this paper is

\begin{equation}
\label{equation2}
V = A(\psi^4 + (\frac{4}{3}k - 2) \psi^3 + (1 - 2k)\psi^2)
\end{equation}

where \(A\) is a constant. The reason for using such a form becomes clear with Figure \ref{figure6}. When \(T<T_{c}\), \(k>0\), and the solid phase is more stable than the solution phase as seen in Figure \ref{figure6}(a). When \(T>T_{c}\), the solution phase is more stable than the solid phase as seen in Figure \ref{figure6} (c).

In the experiment, dendritic growth of PEO outside the pattern exhibits four-fold anisotropy. Since anisotropy depends on crystal structure and other material constants, it is reasonable to assume that the pillars do not change the intrinsic anisotropy of the polymer. Rather, as is shown by the simulation results, the apparent eight fold anisotropy arises out of the inherent four-fold anisotropy of PEO, plus the constraints imposed by the pillars. Therefore, we introduce (intrinsic) four-fold anisotropy into the system by assuming that \(\xi\) depends on the direction of the normal vector at the interface as follows

\begin{equation}
\label{equation3}
\xi = \xi_{o} (1 + \beta \cos(m(\theta)))
\end{equation}

where \(\xi_{o}\) is a constant, \(\beta\) is the strength of the anisotropy, \(m\) is the mode of anisotropy (in our case, \(m=4\)) and \(\theta\) is the angle made by the normal vector at the interface with the positive \(x\) direction.

The equation governing the evolution of \(\psi\) is an Allen-Cahn type of equation, namely,

\begin{equation}
\label{equation4}
\eta \frac{\partial \psi}{\partial t} = - \frac{\delta F}{\delta \psi}
\end{equation}

where \(\eta\) is a positive constant. After simplifications, equation \ref{equation4} can be written as

\begin{multline}
\label{equation5}
\eta \frac{\partial \psi}{\partial t} = - \frac{\partial}{\partial x}(\xi \frac{d\xi}{d\theta} \frac{\partial \psi}{\partial y}) + \frac{\partial}{\partial y}(\xi \frac{d\xi}{d\theta} \frac{\partial \psi}{\partial x})\\ + \nabla \cdot (\xi^2 \nabla \psi) + 4A\psi (1- \psi)(\psi - \frac{1}{2} + k)
\end{multline}

Temperature is governed by the conduction equation

\begin{equation}
\label{equation6}
C\frac{\partial T}{\partial t} = \kappa \nabla^2 T + P \frac{\partial \psi}{\partial t}
\end{equation}

where \(C\) is the heat capacity per unit volume of the system, \(\kappa\) is the thermal conductivity and \(P\) is the latent heat per unit volume released during phase segregation. In the forthcoming simulations, we set the initial temperature of the system to be 0 and the equilibrium temperature to be 1. Since \(\psi\) varies between 0 (solution phase) and 1 (solid), it can be inferred from equation \ref{equation6} that if \(C/P >\)  1, then the whole system will be solidified (since the average temperature of the system will be less than 1 even after all the latent heat has been released). If \(C/P <\) 1, then \(C/P\) represents the fraction of the domain that will be solidified (since the average temperature will rise to \(\geq\)1, meaning that solidification becomes energetically unfavorable).

\subsection{\label{sec:level2}Numerical Simulations}

Numerical simulations are run on a 2D domain whose dimensions are given in Table A1 in the appendix. A simple finite difference scheme is used to solve equations \ref{equation5} and \ref{equation6}. At every time step, noise of the form \(\alpha \zeta\) is added to the term \(k\), where \(\zeta\) is a random variable on [-1,1] and \(\alpha\) is a constant. This is done to simulate thermal fluctuations and other instabilities, which create non-uniformities giving rise to branches in the growth. Various constants used in the simulation are given in Table A1. As mentioned before, the exact values of microscopic parameters for PEO (or for that matter, any polymer) are not known, so some sample values are used to show how phase field modeling yields qualitatively similar results to the experiment. Temperature is non-dimensionalized so that the (non-dimensional) critical temperature is 1, and the initial temperature is zero.

Two types of simulations are carried out. The first type simulates growth on an unpatterned surface, while the second has a pattern similar to that in the experiment. The pillars in the pattern are maintained at an `effective temperature' of 5, so that PEO avoids the pillars by virtue of minimizing its free energy. Empirical evidence clearly suggests that the dendritic growth pattern avoids the patterned pillars. This experimental fact can be incorporated in the model by maintaining the pillars at this high value of effective temperature. To maintain the pillars at an effective temperature of 5 and avoid leakage, an adiabatic boundary condition is imposed between the pillars and the rest of the domain while solving equation \ref{equation6}. We assume that there is no heat loss or heat influx at the boundaries and therefore impose adiabatic boundary conditions for temperature. Temperature throughout the domain is initially set to 0 (except for the pillars, whose temperature is fixed to be 5). To prevent the growth from leaving the domain, we impose adiabatic boundary conditions on the order parameter. The results are unaffected even if this choice of boundary conditions is not imposed, however, this choice allows for clarity of presentations of the simulation results. The order parameter is set to be 0 (solution phase) throughout the domain, except a small 3 \(\times\) 3 semi-circular area in the middle of the left boundary wall. This is done to provide an initial seed/nucleation site from which growth can occur. The initial and boundary conditions are summarized in Tables \ref{tab:table2} and \ref{tab:table3}. \\

\begin{table}[h!]	
	\centering
	\caption{Boundary Conditions}
	\label{tab:table2}
	\begin{tabular}{ c  c  c }
		
		\hline\hline&&\\[-0.75em]
		
		\hspace{1 cm}Variable\hspace{0.5 cm} & \hspace{0.5 cm}Boundaries\hspace{0.5 cm}	& \hspace{0.5 cm}Condition\hspace{1 cm}  \\ 
		
		\hline&&\\
		
		\(\psi\) & x=0,x=9      &\(\frac{\partial \psi}{\partial y}=0 \) \\ &&\\
		\(\psi\) & y=0,y=9      & \(\frac{\partial \psi}{\partial x}=0\) \\ &&\\
		\(T\)    & x=0,x=9      & \(\frac{\partial T}{\partial y}=0\)\\ &&\\
		\(T\)    & y=0,y=9      & \(\frac{\partial T}{\partial x}=0\)\vspace{0.2cm}\\
		
		\hline
		
	\end{tabular}
	
\end{table}
\begin{table}[h!]	
	\centering
	\caption{Initial Conditions (\(D\) refers to the entire domain)}
	\label{tab:table3}
	\begin{tabular}{ c  c  c }
		
		\hline\hline&&\\[-1em]
		
		\hspace{0.3 cm}Variable\hspace{0.2 cm} & \hspace{0.2 cm}Region\hspace{0.2 cm}	& \hspace{0.2 cm}Value\hspace{0.4 cm}  \\ 

		\hline&&\\

		\(\psi\) & \(\Omega=(x,y) \in D \mid (x-150)^2+y^2<9\)      & 1 \\ &&\\
		\(\psi\) & \(D-\Omega\)      & 0 \\ &&\\
		\(T\)    & \(D\)   & 0\vspace{0.2cm}\\ \hline

\hline
		
	\end{tabular}
	
\end{table}

\vspace{1cm}
\begin{figure*}
	\centering
	\includegraphics[width=1\textwidth]{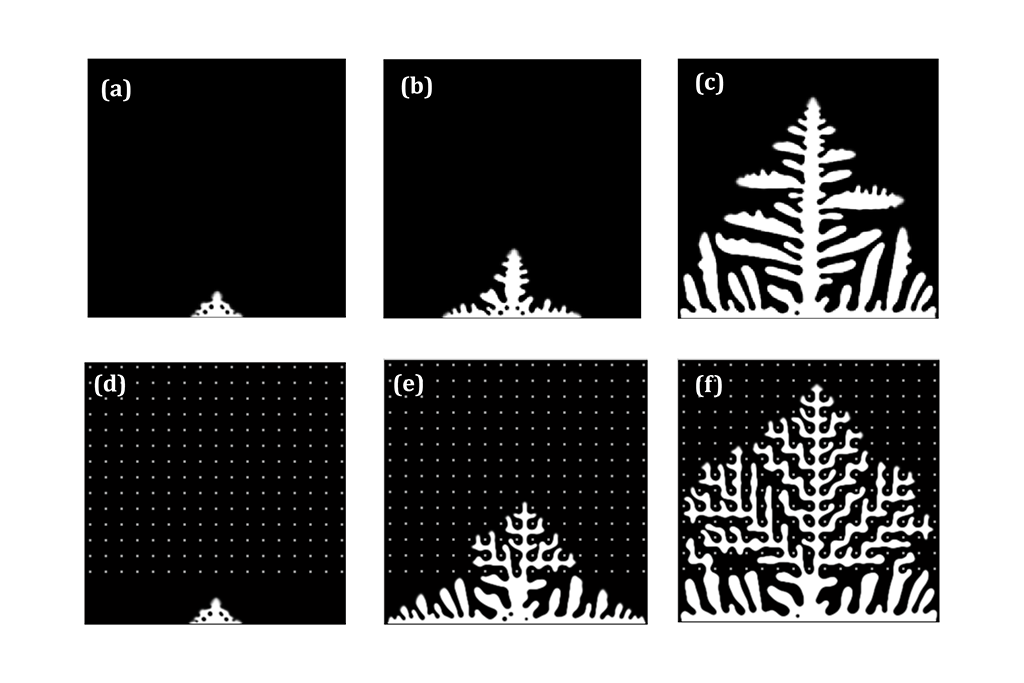}
	\caption{Simulation results on a (i) bare substrate for the following time durations (a) 0.1 s, (b) 0.3 s, (c) 1.0 s, and (ii) patterned substrate for the following time durations (d) 0.1 s, (e) 0.6 s, (f) 1.2 s.}
	\label{figure7}
\end{figure*}

\subsection{Results and Discussion}

\subsubsection{Comparing simulations of polymer growth on a bare substrate and patterned substrate with experimental results}

Figure 3(a)-(c) shows simulation results at different times for an bare substrate. Note that the last image does not represent the end of growth. It is seen that this qualitatively resembles growth outside the patterned region in the experiments (as shown in Figure 1). The growth is anisotropic, with the principal directions aligned with the vertical and horizontal axes, and is also self avoiding. The reason for self avoidance is that whenever the order parameter in a region increases (i.e., when solidification occurs), latent heat is released into the surrounding region, thus decreasing supercooling and reducing the driving force for solidification.

Figure 3(d)-(f) shows simulation results at different times for a patterned substrate. Again, the last image does not denote conclusion of the growth. It is seen that the result qualitatively matches with experiment. The growth avoids the pillars and itself, while giving rise to inclined branches, which makes it seem like it exhibits eight-fold anisotropy \cite{footnote1}.

\subsubsection{Role of pillar separation}

In the simulation on a patterned substrate in the previous section, the pillar separation was such that eight fold anisotropy was observed. In this section it is shown that changing pillar separation changes the overall anisotropy of the final growth structure. 

\vspace{1cm}
\begin{figure*}
	\centering
	\includegraphics[width=0.9\textwidth]{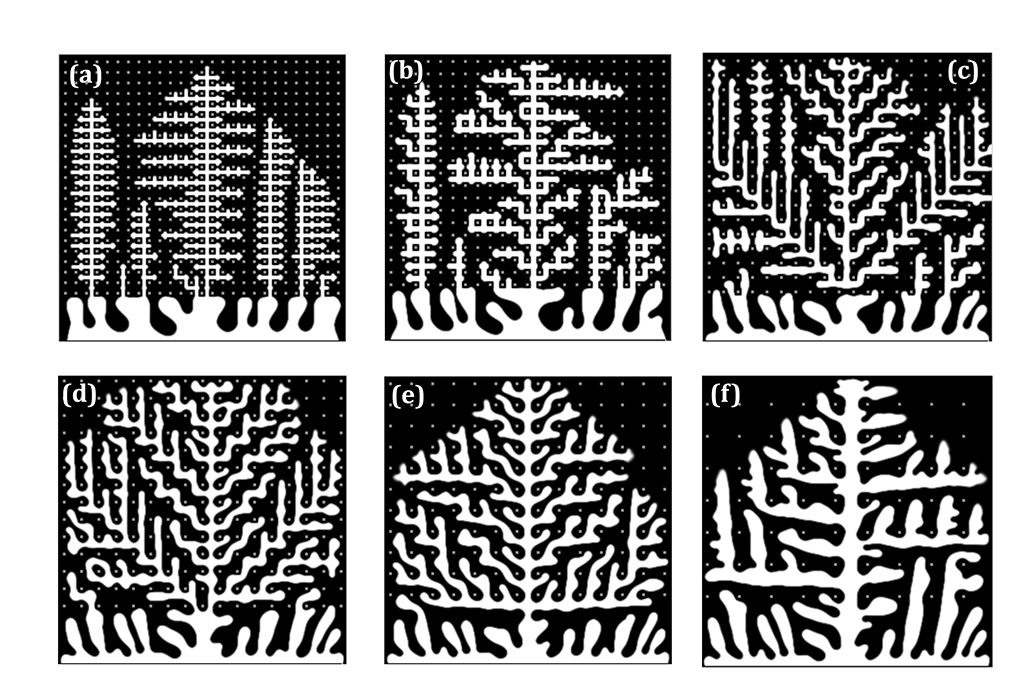}
	\caption{Simulation results on substrates with increasing pillar separation for the following time durations (a) 8.0 s, (b) 5.0 s, (c) 2.2 s, (d) 1.6 s, (e) 1.4 s, (f) 1.4 s.}
	\label{figure8}
\end{figure*}

Figure 4 shows growth patterns on substrates with different pillar separations. At low separations, it is seen that four-fold anisotropy is dominant (Figures 4(a) and (b)). The reason is that the pillars are so close together (in relation to the thermal diffusion length) that it is energetically infeasible for inclined branches to grow. As pillar separation is increased, we start to observe eight fold anisotropy (Figures 4(c) and (d)). If the pillar separation is increased even further, eight-fold anisotropy starts to decrease (Figure 4(e)) until the pillars are so wide apart that they do not have a substantial effect on the original anisotropy of the polymer (Figure 4(f)). 

An interesting observation is that the rate of growth is slower when the pillars are closer (as evidenced by the time duration corresponding to the growth in Figure 4). The reason for this is that when pillars are closer, it is relatively harder for the polymer to grow in between the pillars because of its surface tension which is kept constant in all simulations.

\subsubsection{Role of polymer anisotropy}	

In the previous section, it was observed that pillar separation has a significant effect on the anisotropy of the growth. However, the polymer had an intrinsic four-fold anisotropy to begin with. In this section, we observe the role of intrinsic anisotropy of the polymer on the overall growth structure.

Figure 5(a) shows the growth pattern for a polymer with four-fold anisotropy, and Figure 5(b) shows the growth pattern for an isotropic polymer. While there are differences in the growth structure, it can be seen that both cases exhibit eight-fold anisotropy. This means that the pattern and anisotropy of pillars, rather than the polymer, decides the anisotropy of the final structure. To test this hypothesis further, we run simulations on a substrate with a hexagonal pillar pattern.

Figure 5(c) shows the growth pattern for a polymer with four-fold anisotropy on a hexagonal lattice, and Figure 5(d) shows the growth pattern for an isotropic polymer on a hexagonal lattice. Again, while there are differences in the growth structure, we see that both cases exhibit eight-fold anisotropy. This means that the anisotropy of pillars has a dominating effect in deciding the anisotropy of the final structure.

\section{CONCLUSIONS}

This paper presented the experimental results involving dendritic growth of PEO on a patterned Si/SiO\(_2\) substrate created using electron beam lithography. It was observed that dendritic growth structures of PEO on Si/SiO\(_2\) substrates display self-avoidance, and avoid the lithographically cross-linked PEO pillars in the case of a patterned substrate. On a bare substrate, it was seen that PEO exhibits four-fold anisotropy, whereas in the patterned substrate, it is found to exhibit eight-fold anisotropy.\\

It was discussed that the phase segregation of PEO is temperature controlled, with quite a low value for the critical temperature (~ 66 \(^o\) C) when the weight fraction of PEO is high, as may typically be found in interaction of PEO with surface water. This resulted in temperature controlled dissolution and segregation, during which dendritic growth occurred. Further, these observations were explained based on the principles of free energy minimization, with the main driving force being temperature driven phase change. Intrinsic four-fold anisotropy was introduced into the model to account for the four-fold anisotropy of PEO. The pillars on the patterned substrate were modeled as regions having high temperature, which we called the `effective temperature'. This made growth in those regions unfavorable, which explains the key experimental observations.. In reality, the 'effective temperature' could also have been electrostatic repulsion or some other forces at play.\\

Numerical simulations using phase field modeling showed that the theory yields results that qualitatively match experiments. Since the exact values of microscopic parameters of any polymer (including well studied systems like PEO) are not known, sample values were used to qualitatively compare the theoretical results and experimental observations. Numerical simulations also revealed that changing pillar separation affects the anisotropy of the system, and that the anisotropy of the final structure depended strongly on the anisotropy of the pattern rather than the polymer's intrinsic anisotropy. The approach used to model the experiment was very general and is applicable to any system involving temperature driven phase change and anisotropic growth.\\

Scope for further research includes determining precise microscopic parameters involved in the system described in the experiment. One could also model the pillars in a different way, for example, using electrostatic forces. There is also the possibility of studying the effect other pattern configurations on the dendritic growth patterns of polymers on substrates. Such research might help better understand pattern formation in thin films, where the role of the substrate is often ignored. Further, predictive capabilities of numerical simulations may be used to uncover behavior in systems that are difficult to fabricate for experiment. 

\vspace{1cm}
\begin{figure*}
	\centering
	\includegraphics[width=1.0\textwidth]{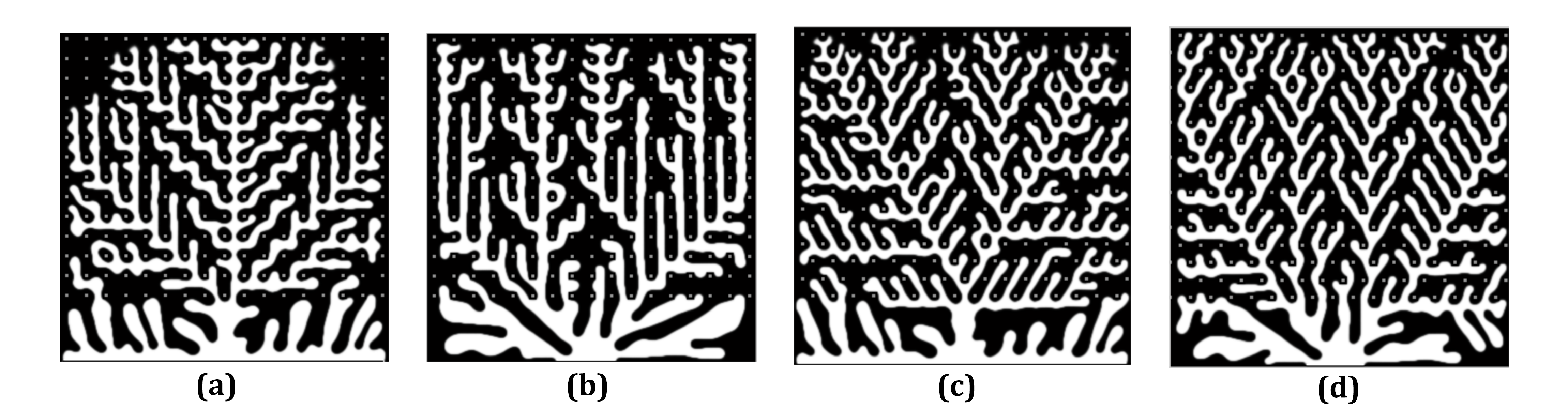}
	\caption{Simulation results on a (i) square patterned substrate with (a) a polymer of four-fold intrinsic anisotropy, (b) a polymer with no anisotropy, (ii) hexagonal patterned substrate with (c) a polymer of four-fold intrinsic anisotropy, (d) a polymer with no anisotropy.}	
	\label{figure9}
\end{figure*}

\setcounter{table}{0}
\renewcommand{\thetable}{A\arabic{table}}

\appendix

\section{\label{sec:level1}Numerical constants used in the simulations}

Table AI shows the values of numerical constants used in the numerical simulations.

\begin{table}[h!]	
	\centering
	\caption{Constants used in the simulation}
	\label{tab:table1}
	\begin{tabular}{ c  c }
		
		\hline\hline&\\
		
		\hspace{1.5 cm}\textbf{Constant}\hspace{1.5 cm}                            &\hspace{1.5 cm}\textbf{ Value}\hspace{1.5 cm}  \vspace{0.2cm}\\ 
		
		\hline&\\
		
		Simulation Domain Size              & 9 x 9      \\&\\[-0.5em]
		Mesh Constants                      & 300 X 300   \\&\\[-0.5em] 
		\(\Delta x, \Delta y\)              & 0.03   \\ &\\[-0.5em]
		\(\Delta t\)                        & 0.00002   \\ &\\[-0.5em]
		\(\mu\)                             & 0.0.9   \\ &\\[-0.5em]
		\(\lambda\)                         & 10   \\ &\\[-0.5em]
		\(A\)                               & 0.25   \\ &\\[-0.5em]
		\(\xi_{o}\)                         & 0.01   \\ &\\[-0.5em]
		\(\beta\)                           & 0.02  \\ &\\[-0.5em]
		\(m\)                               & 4 \\ &\\[-0.5em]
		\(\eta\)                            & 0.0003    \\ &\\[-0.5em]
		\(C\)                               & 1 \\ &\\[-0.5em]
		\(\kappa\)                          & 1 \\ &\\[-0.5em]
		\(P\)                               & 1 \\ &\\[-0.5em]
		\(T_{c}\)                           & 1 \\&\\
		
		\hline
		
	\end{tabular}
	
\end{table}


\begin{thebibliography}{9}

	\bibitem{paper8} 
	E. Ben-Jacob, 
	Contemporary Physics \textbf{34}, 247 (1997).
	
	\bibitem{paper10} 
	A. J. Koch and H. Meinhardt,
	Rev. Mod. Phy. \textbf{66}, 1481 (1994).
	
	\bibitem{paper11} 
	A. Gierer, Prog. Biophys. Molec. Biol. \textbf{37}, 1 (1981).
	
	\bibitem{paper9} 
	U. Nakaya, 
	\textit{Snow Crystals: Natural and Artificial} 
	(Harvard University Press, Cambridge, 1954).	
	
	\bibitem{paper1} 
	M. Cross and H. Greenside, 
	\textit{Pattern Formation and Dynamics in Nonequilibrium Systems} (Cambridge University Press, New York 2009).

	\bibitem{paper12} 
	M. E. Glicksman, 
	\textit{Handbook of Crystal Growth (Second Edition)} (Elsevier, Boston, 2015), Vol. IB, Chap. 16.

	\bibitem{paper13} 
	M. A. Zaeem and L.M. Hogan, 
	in \textit{Reference Module in Materials Science and Materials Engineering}, (Elsevier, 2017).
	
	\bibitem{paper14} 
	C. C. Hoogenraad and C. Wierenga, in 
	\textit{Reference Module in Biomedical Sciences, from Encyclopedia of the Neurological Sciences (Second Edition)} (Academic Press, Oxford, 2014), p. 456.
	
	\bibitem{paper15} 
	Jens Feder, 
	\textit{Fractals} 
	(Springer, Boston, 1988).
	
	\bibitem{paper16} 
	H. W. Leung, in
	\textit{Reference Module in Biomedical Sciences, from Encyclopedia of Toxicology (Third Edition)} (Academic Press, Oxford, 2014), p. 1043.
	
	\bibitem{paper17} 
	K. Hillier, in
	\textit{xPharm: The Comprehensive Pharmacology Reference} 
	(Elsevier, Boston, 2007), p. 1.
	
	\bibitem{israelachvili} 
	J. Israelachvili,
	Proc. Nat. Acad. Sci. U.S.A. \textbf{94}, 8378 (1997).

	\bibitem{devanand} 
	K. Devanand and J. C. Selser, Nature \textbf{343}, 739 ( 1990).

	\bibitem{hammouda} 
	B. Hammouda, D. L. Ho, and S. Kline, 
	Macromolecules \textbf{37}, 6932 (2004).	
		
	\bibitem{shaina} 
	P. R. Shaina and M. Jaiswal, Appl. Phys. Lett. \textbf{105}, 193103 (2014).
	
	\bibitem{paper18} 
	H.-G. Braun, E. Meyer, and M. Wang, 
	\textit{Polymer Crystallization, Observations, Concepts and Interpretations}. 
	(Springer-Verlag, Berlin, 2003), p. 238.
	
	\bibitem{paper19} 
	V. Ferreiro, J. F. Douglas, J. Warren, and A. Karim, Phys. Rev. E \textbf{65}, 051606 (2002) 
	
	\bibitem{paper20} 
	S. Amir, S. A. H. Ali and N. S. Mohamed, Ionics, \textbf{17}, 121 (2011).
	
	\bibitem{paper21} 
	Y. Zhou, X Han, H. Liu, and Y. Hu, Chinese Journal of Chemical Engineering, \textbf{22}, 339 (2014).
	
	\bibitem{paper22} 
	Z. Ma, G. Zhang, X. Zhai, L. Jin, X. Tang, M. Yang, P. Zheng and W. Wang, Polymer
	\textbf{49}, 1629 (2008).
	
	\bibitem{seuring} 
	J. Seuring and S. Agarwal, ACS Macro Lett. \textbf{2} 597 (2013).
	
	\bibitem{bekiranov} 
	S. Bekiranov, R. Bruinsma, and P. Pincus, Phys. Rev. E \textbf{55}, 577 (1997).

	\bibitem{paper2} 
	G. J. Fix, 
	\textit{Free Boundary Problems: Theory and Applications} (Pitman, Boston, 1983), p. 580.
	
	\bibitem{paper3} 
	 J. S. Langer,
	\textit{Directions in Condensed Matter Physics}. 
	(World Scientific, Singapore, 1986), p. 165.	
	
	\bibitem{paper4} 
	A. Karma and W.-J. Rappel, Phys. Rev. E \textbf{57}, 4323 (1998)
	
	\bibitem{paper5} 
	W. J. Boettinger, J. A. Warren, C. Beckermann, A. Karma, Annual Review of Materials Research \textbf{32}, 163 (2002). 
	
	\bibitem{paper6} 
	L.-Q. Chen , Annual Review of Materials Research \textbf{32}, 113 (2002). 
	
	\bibitem{paper7} 
	R. Kobayashi, Physica D \textbf{63}, 410 (1993).
	
	\bibitem{footnote1}
	When we say eight-fold anisotropy, we refer to the anisotropy on the scale of the growth. If magnified, the growth still exhibits four-fold anisotropy between pillars, but the larger structure clearly exhibits eight fold anisotropy.
	
\end{thebibliography}
\end{document}